\documentclass[a4paper,10pt,twoside]{cpc-hepnp}

\usepackage{multicol}
\usepackage{graphicx}
\usepackage{booktabs}
\usepackage{amssymb,bm,mathrsfs,bbm,amscd}
\usepackage[tbtags]{amsmath}
\usepackage{lastpage}
\usepackage{CJK}

\begin{document}

\begin{CJK*}{GBK}{song}

\fancyhead[c]{\small Submitted to 'Chinese Physics C'} \fancyfoot[C]{\small 010201-\thepage}

\footnotetext[0]{Received 12 June 2013}

\title{Acceleration Modules in Linear Induction Accelerators}

\author{%
      Shaoheng Wang$^{1)}$\email{wang@jlab.org}%
\quad Jianjun Deng
}
\maketitle

\address{%
Institute of Fluid Physics, China Academy of Engineering Physics, Mianyang, 621900 China}

\footnotetext[0]{Currently working at JLAB, USA}

\begin{abstract}
Linear Induction Accelerator (LIA) is a unique type of accelerator, which is capable to accelerate kilo-Ampere charged particle current to tens of MeV energy. The present development of LIA in MHz bursting mode and successful application into synchrotron broaden LIA's usage scope. Although transformer model is widely used to explain the acceleration mechanism of LIAs, it is not appropriate to consider the induction electric field as the field which accelerates charged particles for many modern LIAs. Authors examined the transition of the magnetic cores' functions during LIA acceleration modules' evolution, distinguished transformer type and transmission line type LIA acceleration modules, and re-considered several related issues based on transmission line type LIA acceleration module. The clarified understanding should be helpful in the further development and design of the LIA acceleration modules.
\end{abstract}



\begin{multicols}{2}

\section{Introduction}

To obtain particle beams of different characteristics in lab, many types of accelerators are invented. Of them, Linear Induction Accelerator (LIA) is a special type of accelerator which can accelerate pulsed, high intensity electron or ion beams to several tens of MeV energy~\cite{Christofilos,Kapetanakos}. Although this energy level is pretty low compared to other types of accelerators, like synchrotrons, RF Linacs, the beam current in LIAs can be as large as one hundred kilo-Amperes~\cite{LIA-30}. LIAs have many special applications: radiography, high power microwave production, FEL, heavy-ion inertial fusion, etc. And the successful application of LIA acceleration technique into synchrotron~\cite{Takayama} even further broadens its potential usages.

The key component of a LIA is its acceleration module, which has a high permeability magnetic core. LIA acceleration modules can be easily stacked in series to accelerate charged particles to high energy, but, this type of architecture also limits the LIA's acceleration gradient. LIA acceleration modules are low-impedance, non-resonant devices which enables LIAs to accelerate kilo-Ampere charged particle current without severe instability problems.

The first LIA, ASTRON-I, was built in 1963, and transformer model is the most widespread model to introduce the acceleration concept in LIA, which attributes the source of the acceleration voltage as the integral of the induced field of the magnetic flux variation through the LIA magnetic core~\cite{Kapetanakos}. The transmission line model was brought about by Keefe in 1981~\cite{Keefe}. In this model, the magnetic core and its housing is considered as a transmission line with one end shorted. The acceleration voltage is attributed to the voltage difference between the inner and outer conductor of the power supply transmission line. The high voltage pulse passes through the acceleration gap region and travels into the core housing. When the high voltage pulse is reflected from the shorted end of the core housing transmission line and travels back to the gap region, the gap acceleration voltage collapses. Although transmission line model is used in LIA acceleration module design, some important characters of LIA modules, like voltage-seconds, beam loading effect and voltage droop are still explained according to transformer model. Smith and others explicitly denied the acceleration voltage to be the magnetic core induced voltage~\cite{Smith, Mascureau}. But, it is not fair to curtly say the transformer model is wrong and some confusion does exist in understanding the LIA acceleration module.

LIA technology is being exploited to provide new features and higher performance, like providing longer pulse, working under MHz busting mode. Authors feel it is necessary to clarify some unclear points for further developing LIA technology. We re-examined LIA acceleration structures and models, discussed several important issues, tried to provide a more accurate understanding of LIA acceleration module.

\section{Induction Cavity and Other type Accelerating Cavities}
\subsection{Induction Cavity and RF Cavity}
If we focus on power flow route in accelerators, we can see that both RF and induction linear accelerators are composed of many power transmission lines, which are all driven in parallel and feed power in to cavities where charged particles are accelerated while passes through them in series.

In the RF accelerators, isolation between the cavities is achieved by connecting them with a beam pipe of which the cutoff frequency is below the cavity operating frequency. RF cavity is a resonant structure with a high quality factor $Q$, and is able to contain a very high field which give it a high acceleration gradient. But both generator current and beam current excite electric-magnetic field in the resonant cavity. The disadvantage of resonant structures is that a cavity is never simply resonant with a single mode and the wake functions of the charged particle beam have Fourier components which feed energy into all available modes. Some of these modes act back to the beam and cause beam instabilities. Since the higher the beam intensity, the stronger the wake fields, this sets an upper limit to the total charge that can be accelerated.

In the LIA, isolation is achieved with the magnetic core of high permeability. An induction cavity is designed to be non-resonant, the low $Q$ property makes it store neither the drive fields nor the beam's wake fields. This dramatically increases the practical operating current. The penalty accompanying these advantages is centered on the fact that induction accelerators are incapable of efficiently accelerating very low current beams.

\subsection{Cavities with Ferrite Cores}
\subsubsection{Ferrite Cores}
In operating induction linacs that provide short pulses ($\sim 50$ ns), the induction module is filled with ferrimagnetic material, ferrite; for long pulses ( 50 ns $\leq \tau_p \leq$ several $\mu $s ) ferromagnetic materials or Metglas~\cite{Rutkowski} are used. This paper focuses on cavities with ferrite cores when we are dealing with transmission line model.

Ferrite is a ceramic like type of material. Ferrites exhibit a form of magnetism called ferrimagnetism, which is different from the ferromagnetism of such materials as iron, cobalt, and nickel. Two crystalline sub-lattices exist in ferrites. A partial cancelation of the magnetic field results, and the ferrite is left with an overall magnetic field that is less strong than that of a ferromagnetic material. The saturate magnetic inductions of NiMn ferrite usually fall in the range of 0.2-0.4 T~\cite{IGardner}.

Ferrite is a very convenient choice of the magnetic material due to its very high resistivity. This resistivity ensures a skin depth of the order of many tens of cm for 50 ns pulse length, which in turn assures efficient utilization of the transection area of the magnetic cores for such pulses.

\subsubsection{Resonant and Non-resonant Cavities with Ferrite Cores}
There are two types of acceleration cavities using large volume of ferrites. They are  Resonant Ferrite Core Cavities (RFCC), referred in this paper, used in synchrotrons and Non-Resonant Ferrite Core Cavities (NRFCC), referred in this paper, used in LIAs. Ferrite cores have different roles in these two types of acceleration cavities.

RFCC are used in synchrotron RF systems with ramping frequencies. In synchrotrons, particles circulate in the beam pipe ring for many thousands of turns and get accelerated when passing through the cavities in each turn. Revolution frequency of particles ramps up while the their energies increase. The resonant frequency of accelerating cavity needs to keep up with particles' revolution frequency. RFCC uses shorted, ferrite loaded coaxial transmission lines as inductances to resonate with the accelerating gap capacitance. There is a DC bias current around the ferrite core in RFCC. The circuit is kept in tune with the required frequency by changing the bias current which in turn changes the permeability of the ferrite core and the resonance frequency, as indicated in Fig.(\ref{Hysteresis}.a). Only a small portion of the saturate Hysteresis loop is used at any moment of the ramping. In this way, the RFCC realize a several MHz or tens of MHz tuning range of the resonant frequency. And the Q value of RFCC ranges from several hundreds to several thousands.
\begin{center}
\includegraphics[width=8cm, angle=0]{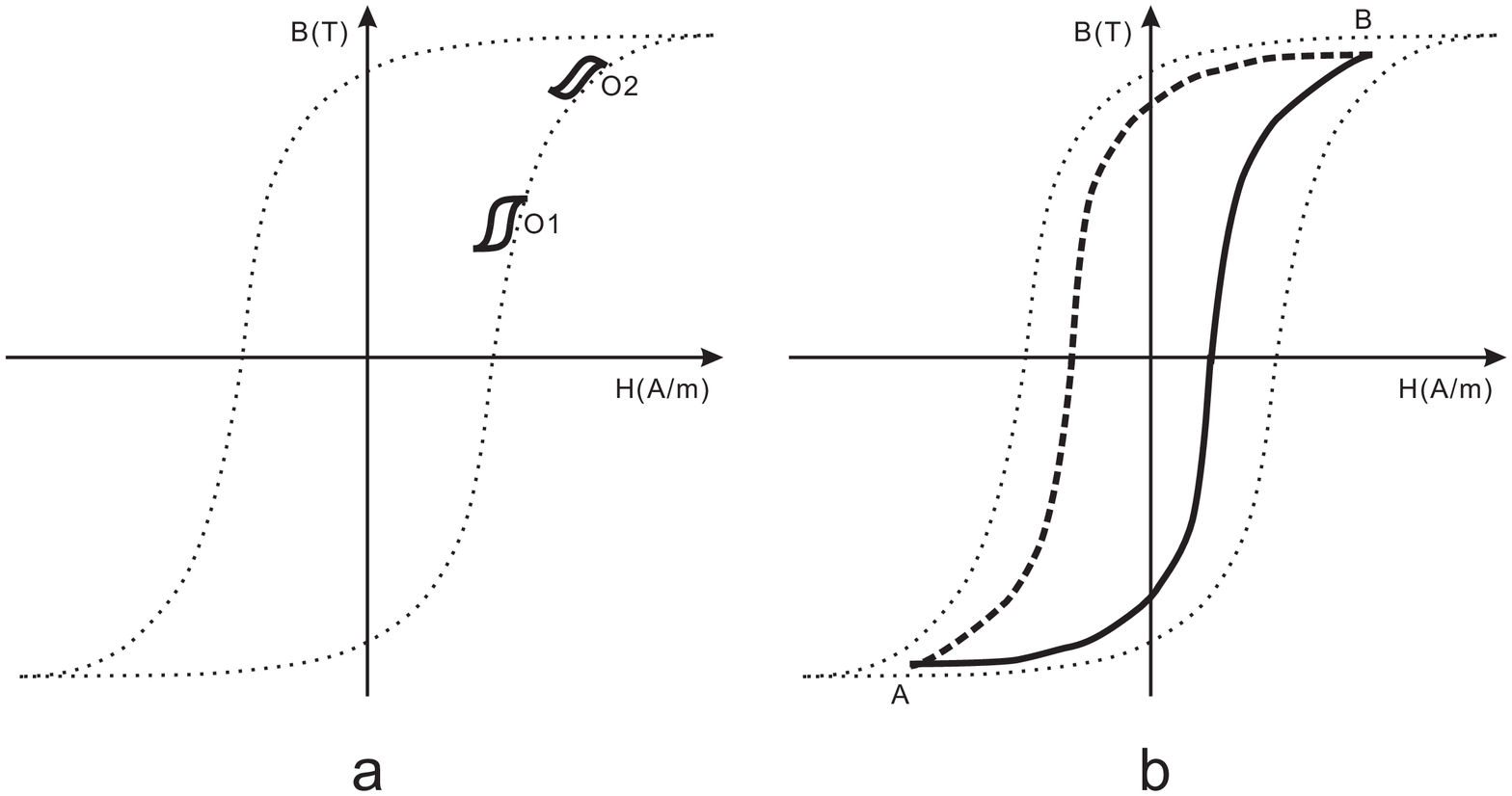}
\figcaption{\label{Hysteresis}Different utilization of the Hysteresis loop in RFCC (graph a) and NRFCC (graph b). }
\end{center}

While, NRFCC in LIA is a non-resonant structure, used in a pulsed working mode. For each pulse, B-H variation of the ferrite cores covers most of the saturate Hysteresis loop to obtain the maximum $\Delta B$, except in the case of bursting mode, where Hysteresis loop range is distributed among several pulses.

In both RFCC and NRFCC, the ferrites allow the size of the cavities to be small.

\section{Development of Induction Cavities and the Models}
\subsection{Transformer Model and Astron}
Astron is the first LIA, and the transformer model of the acceleration mechanism in LIA was clearly stated with Astron~\cite{Christofilos}. For the purpose of comparison with later developed LIA cavities and transmission line model, we will briefly restate the Astron acceleration module and transformer model here.
\begin{center}
\includegraphics[width=5cm, angle=0]{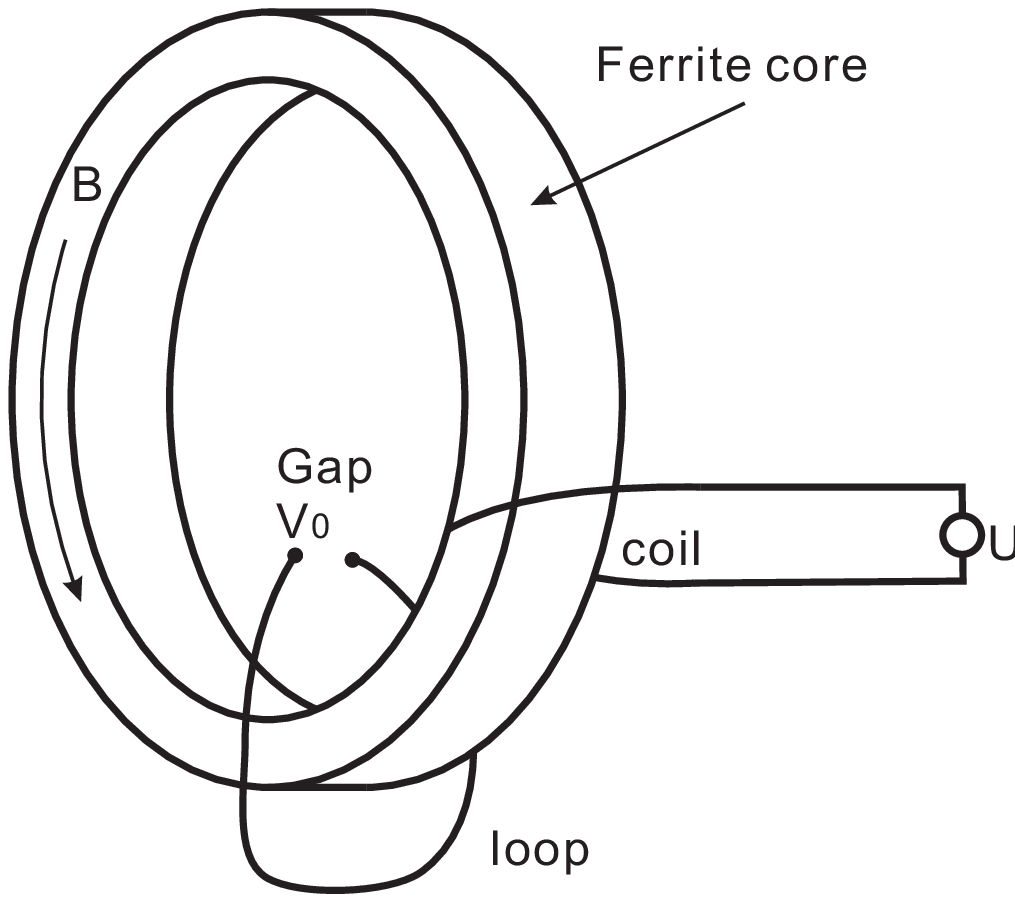}
\figcaption{\label{Fig:TransformerModel}Diagrams of transformer model. }
\end{center}
The geometry of the accelerating structure is shown diagrammatically in Fig.~\ref{Fig:TransformerModel}. A toroidal ring of magnetic material, the core,  surrounds of the acceleration column (not shown in the figure), and the change of flux in the magnetic core induces an axial electric field, which is the acceleration field and appears at the gap of the secondary loop,
\begin{equation}\label{eq:VS0}
    V_0 = \oint \vec{E}\cdot d\vec{l}
        =-\int_s \dfrac{\partial \vec{B}}{\partial t}\cdot d\vec{S}.
\end{equation}
where the $\vec{E}$ is integrated along the secondary loop and $\dfrac{\partial \vec{B}}{\partial t}$ is integrated over the transection area of the magnetic core.
Each core in ASTRON is threaded by three primary straps symmetrically placed. The case of each core forms the secondary coil. Electrodes are connected to the plate of the case. Each core contributes an average of 12 keV to the beam. In the electron gun of ASTRON, the required field gradient is obtained by adjustment of electrode spacing and the number of cores between successive gaps.

The requirement for energy uniformity during the beam pulse makes it necessary to make some provision in the pulsing system to maintain a constant $\dfrac{dB}{dt}$ during the useful part of the pulse. This was done by providing a current ramp in the primary pulse. A specially designed pulse shaper circuit was used to produce the required current ramp. The primary and secondary voltage pulses are copied here in Fig.~\ref{Fig:AstronVoltages}
\begin{center}
\includegraphics[width=8cm, angle=0]{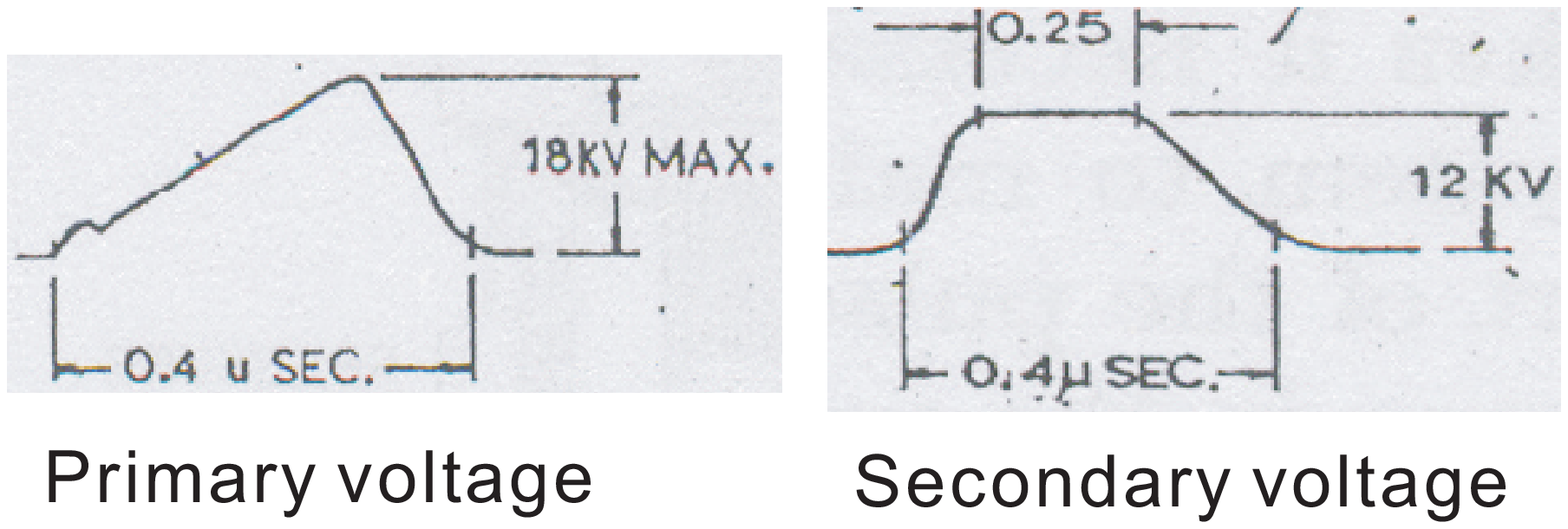}
\figcaption{\label{Fig:AstronVoltages}Astron's primary and secondary voltages.}
\end{center}
The transformer model perfectly explains the acceleration mechanism in Astron's experiments.

\subsection{Later developed Induction Cavities and Transmission Line Model }

In Astron's case, the dependence of constant voltage on $\dfrac{dB}{dt}$ is clearly explained and demonstrated. But, the relationship of the flux swing $\Delta$B to H and the drive current $I$ is complicated for the magnetic core, being nonlinear and time dependent for the pulse durations of interest. And the insulation problem in the assembly of Astron required considerable efforts during the development of Astron. During the journey to higher accelerating gradient, more magnetic cores were placed in single case while magnetic isolation concept is developed. And the structure of the so-called LIA acceleration module has subtly changed.

Figure~\ref{Fig:LIA_Modules} shows the diagram of a typical later developed LIA acceleration modules. Although the inner conductor of the power feed-in cable is connected with the outer conductor through the conductor surrounding the cores, it is inductive isolated because of the large core inductance when the high voltage pulse is applied. Hence, the outer conductor of the core housing won't see the high voltage traveling along the inner conductor, the high voltage appears only inside the core housing. The benefits may include suppression of electromagnetic interference to nearby systems, avoiding electrical breakdown, the ability to make parallel connections upstream in the driver circuits. The inductive fields cancel electrostatic fields over the surrounding conductor, while the beam still senses the acceleration field at the gap. This process is called inductive isolation.
\begin{center}
\includegraphics[width=8cm, angle=0]{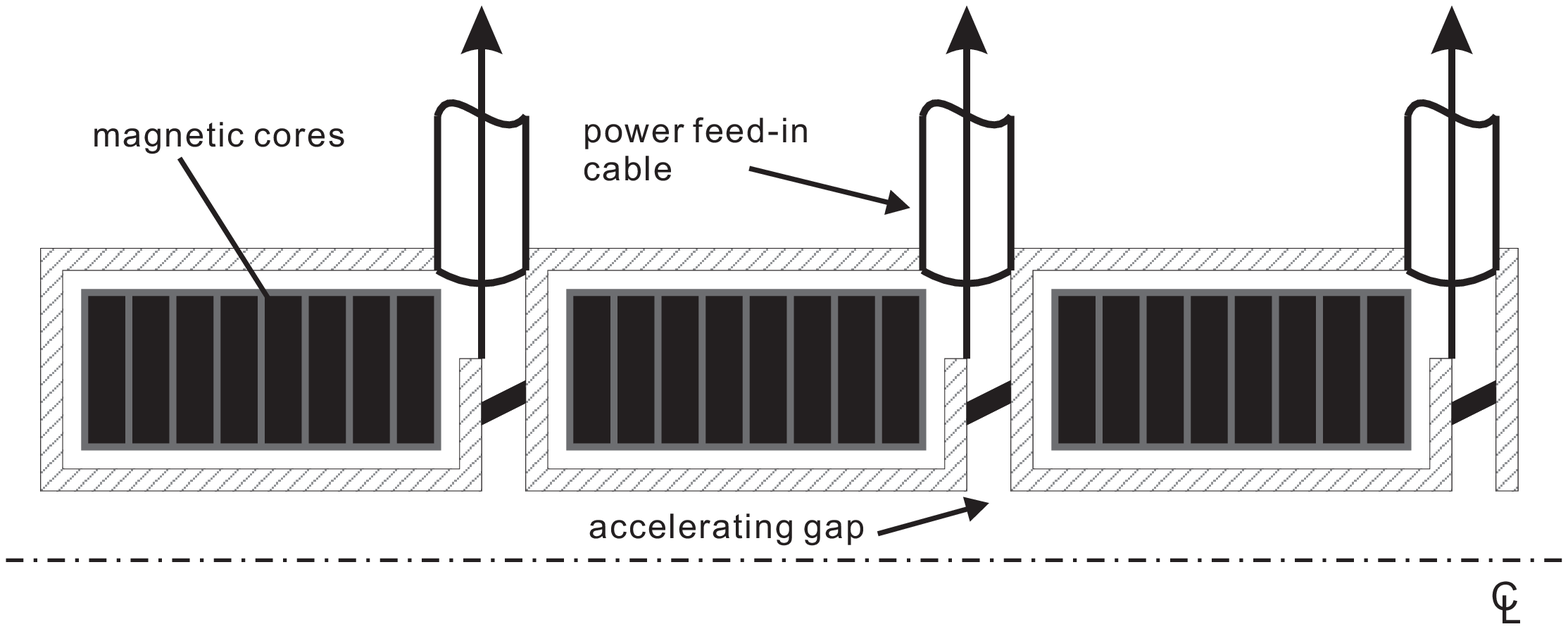}
\figcaption{\label{Fig:LIA_Modules}Diagram of later developed LIA cavities.}
\end{center}

In such kind of structure, transformer model is challenged for the explanation of the acceleration mechanism. In Astron and transformer model, the acceleration voltage is obtained through magnetic coupling and appears in the gap of the secondary loop. In the later developed LIA cavities, the coupling from the power system to the acceleration gap is by hard connection, and the view that $\dfrac{dB}{dt}$ causes the acceleration seems inappropriate. Whether there is a hard connection from the high voltage conductor to the gap is the key difference between the two acceleration structures. In the later developed LIA cavities, the high voltage conductor entering the vacuum bore must connect back to the outer ground surface of the cell to avoid the charged particle beam seeing the decelerating field. In order to keep the loss current that flows from high voltage to ground through this connection path to a level low enough that it does not drop the high voltage drive pulse down unacceptably, the inner conductor is surrounded by ferrite cores.

There is always an image current accompany the beam current in the beam pipe and forms a current loop with the beam current. At the accelerating gap, the image current flows into the power supply cable. Hence, the power supply current is divided into a loading current flowing into the gap, and an magnetization current flowing around the magnetic cores. These cores present a large inductance preventing the line from being short-circuited on the electric ground. The loading current equals the beam current. It has to be noticed that the loop formed with the image current and beam current never surrounds the magnetic cores to behave as the secondary loop unlike the transformer model has assumed.

One important parameter of the beam in a LIA is the energy spread, which is determined by the flatness of the acceleration gap voltages. In ASTRON, the input driving voltage pulse is not even a flat-top shape like those in later developed LIA modules. A constant $dB/dt$ was pursued to obtain the best available voltage flatness as observed in Fig.~\ref{Fig:AstronVoltages}. This is because the core of the Astron structure is thin, the inductive impedance of the core is relatively low in the circuit, the magnetizing current is easily controlled by external power supply, hence, constant $dV/dt$ can be used to obtain constant $dI_{drive}/dt$, $dB/dt$ and constant gap voltage in turn.

In the acceleration structure shown in Fig.~\ref{Fig:LIA_Modules}, the electric-magnetic wave process needs to be applied in analyzing the circuit. Magnetic cores and the housing can be viewed as either an large inductance or a short transmission line depending on the electrical length of the core cell compared with the rising time of the high-voltage pulse. When the electrical length of the core cell is shorter, the cores can be viewed as a lump inductance, the magnetizing current loops the cores in a fraction of  the high-voltage pulse length, the magnetic flux density $B$ varies almost uniformly in the full transect when the magnetizing current surrounding the cell varies. The back-EMF around the core cell almost equals the applied high-voltage pulse. The magnetizing current keeps increasing to produce enough back-EMF, this varying magnetizing current results in the gap voltage drooping. When the electrical length of the core cell is longer than the rising time of the high-voltage pulse, it behaves more like a short transmission line, and a constant resistive impedance is seen from the input port, the acceleration gap. The gap voltage in this situation is determined by the high-voltage pulse shape and beam loading variation, hence the flatness of the supplied high voltage pulse is pursued instead of the linearity of $dI_{drive}/dt$ in ASTRON. More about the short transmission line case is analyzed in section~\ref{Sec:Issues}.

The transmission line model was brought forward by Keefe in 1981, see Fig.~\ref{Fig:TLModel}. The power supply cable and the acceleration module are considered as one transmission line filled with different materials. Charged particles are accelerated by the voltage difference between the inner and outer conductor of the transmission line and travel inside the inner conductor. Transmission line model is straight forward in explaining acceleration mechanism in the structure illustrated in Fig.~\ref{Fig:LIA_Modules}.
\begin{center}
\includegraphics[width=5cm, angle=0]{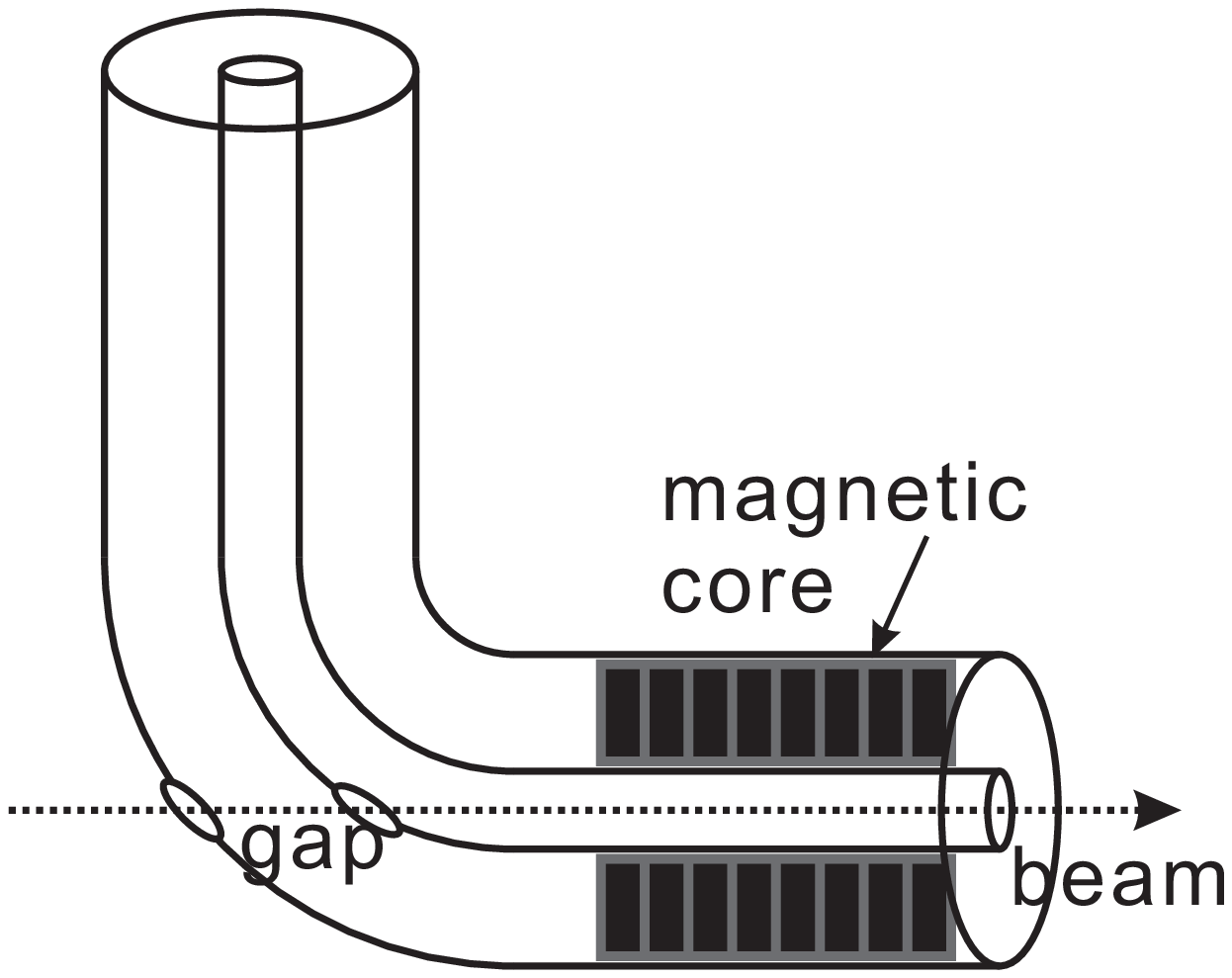}
\figcaption{\label{Fig:TLModel}Diagrams of transmission line model.}
\end{center}

From above discussions, we can see that both transformer model and transmission line model are good in explaining their corresponding structures.

\section{Several Issues about Transmission Line Model}\label{Sec:Issues}

Transmission line type LIA acceleration modules are adopted in many accelerator designs because of their simple and clear description and good gap-voltage-flatness. But several transmission line model related issues still need to be further clarified explicitly, they are discussed below.

\subsection{The Induction Field in LIA Acceleration Modules}

Since we are discussing LIA, the Linear \textbf{Induction} Accelerator, we are obligated to indicate the role of the induction field in both transformer type module and transmission line type module. It is clear that the induction field in transformer type module appears between the electrodes of the secondary loop and functions as the acceleration field.

For a TEM plane wave in a transmission line, like in LIA cases, the phase velocity equals to the speed of light. If we assume $\varepsilon_r$ and $\mu_r$ of our transmission line are constant for all frequency components concerned, then the shape of the pulse will not be distorted. Hence, we can conveniently consider the transportation of the pulse in time domain.

Now, let's consider a pulse with a flat top  and a wave front in a ferrite-filled coaxial transmission line as illustrated in Fig.~\ref{WaveInCoaxial}. The transmission line is shorted at the left end. The incident pulse comes from the right.
\begin{center}
\includegraphics[width=8.5cm, angle=0]{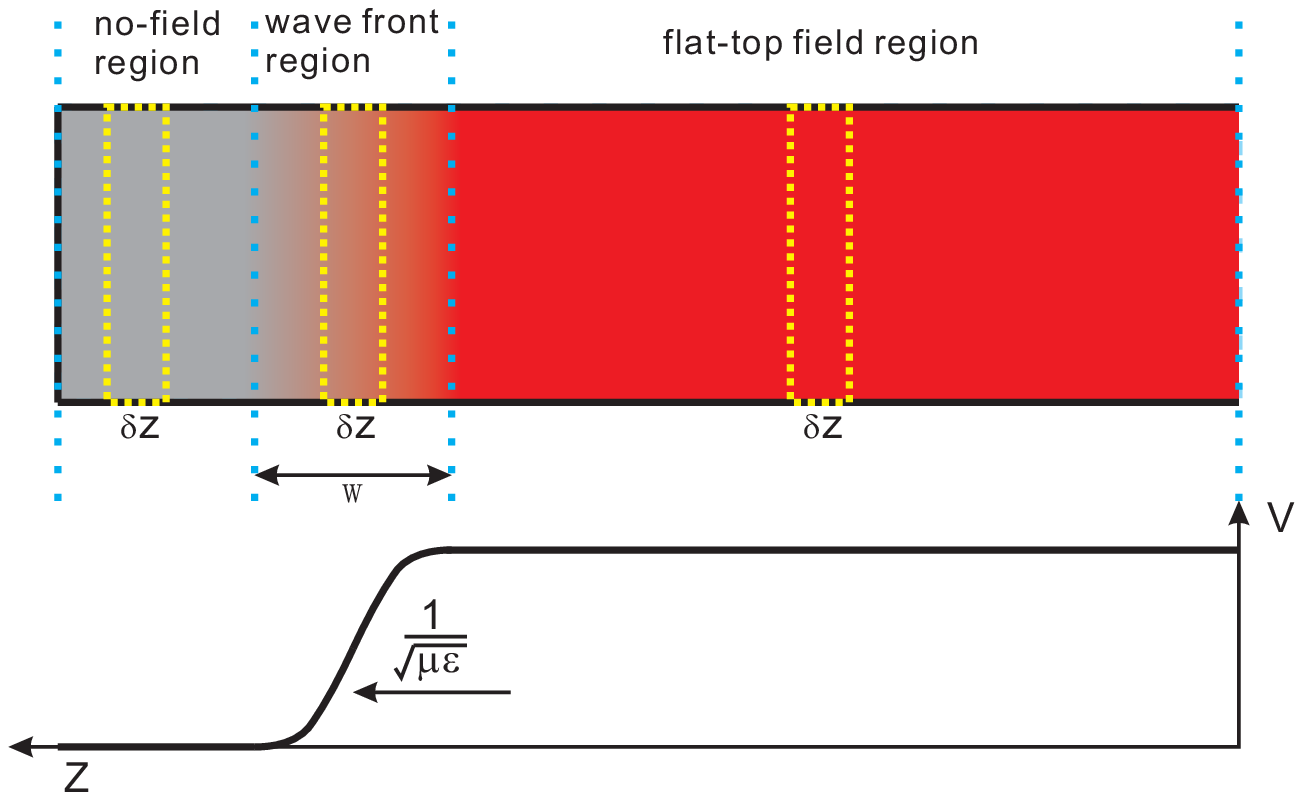}
\figcaption{\label{WaveInCoaxial}TEM wave travels in a coaxial transmission line.}
\end{center}

The wavefront is moving from right to left at a speed of $\dfrac{1}{\sqrt{\mu\varepsilon}}$. At any moment, the EM field distribution in the transmission line can be divided into three regions, the no-field region before the wave front, the wave front region and the flat-top field region. The pulse is long compared with the length of the transmission line, hence the end of the pulse can be ignored at this moment.

When the pulse travels from right to left, the area of the flat-top field region keep increasing constantly. Since the flux density in  the flat-top field region is constant, the flux change rate $\dfrac{d\phi}{dt}$ is constant in the whole area occupied by the transmission line. The increasing of flux $phi$ is resulted from the area increasing of the flat-top field region, instead of increasing of flux density $B$ in the area as in transformer type module. Of course this flux variation produces an induction field.

Next, let's consider three narrow rectangular regions of length $\delta$z in the three regions of the transmission line, as indicated with yellow dotted lines in Fig.~\ref{WaveInCoaxial}. Because no magnetic flux change happens in regions before and after wave front, there is no induction field produced in these two regions. In the moving wave front region, when the wave front sweep through the narrow slice, the magnetic flux inside it keep increasing and the inductive field is produced. The integral of the induction field along the inner conductor through the wave front region is the EMF, which almost cancels the voltage of the incident pulse and provides the isolation between the shorted end and the high voltage input on the right end of the transmission line.

When the electrical length of the structure of the acceleration module is shorter or comparable with the length of the wave front, the wave front region in Fig.~\ref{WaveInCoaxial} extends to cover all the structure, then the structure behaves as an lump inductance.

So, we still have an induction field in transmission line type module, only it's limited inside the core cell, its function is to sustain the high voltage at the acceleration gap.

\subsection{Volt-Seconds}

There is a important design criteria in determining the dimension of the magnetic cores, the Voltage-Second criteria. It's derived from the transformer model and used for every type of LIA module. For a transformer type module, let's rewrite Eq.(\ref{eq:VS0}) here for convenience,
\begin{equation}\label{eq:VS1}
    V_0 =-\oint \vec{E}\cdot d\vec{l}
        =\int_s \dfrac{\partial \vec{B}}{\partial t}\cdot d\vec{S}.
\end{equation}
If we neglect the radial variations of $\vec{B}$ the integration of the Eq.(\ref{eq:VS1}) during the pulse period $\tau$, we have:
\begin{equation}\label{eq:VS2}
    V_0 \tau \approx \Delta B S.
\end{equation}
To get the maximum $\Delta B$, we can preset the cores at the $B_r$ remanence field before each pulse,
\begin{equation}\label{eq:VS3}
    \Delta B_m = -B_r + B_s.
\end{equation}
where $B_s$ is the value of $B$ at saturation. Then, we get the Voltage-Second criteria:
\begin{equation}\label{eq:VS4}
    V_0 \tau \leqslant \Delta B_m S.
\end{equation}

In fact, same result can be derived from transmission line model. In the transmission line model, during the passage of the particle beam through the gap, the high voltage pulse travels inside the transmission line and is reflected at the shorted end, travels back to the acceleration gap. After the reflected high voltage pulse arrives at the accelerating gap, the transmitted wave will spoil the acceleration voltage at the gap. Hence, two times of the high voltage pulse traveling through the transmission line must larger than the beam's length in time. That is
\begin{equation}\label{eq:VSlength1}
    \dfrac{2L}{\dfrac{c}{\sqrt{\varepsilon_f\mu_f}}} \geqslant \tau.
\end{equation}
Rewrite Eq.(\ref{eq:VSlength1}), we get a condition for core cell length $L$:
\begin{equation}\label{eq:VSlength2}
    L \geqslant \dfrac{\tau}{2}\dfrac{c}{\sqrt{\varepsilon_f\mu_f}}.
\end{equation}

In the transmission line model, the flux density difference before and after the wave front is $\Delta B$. As explained in last subsection, the induction field is produced by the increasing of the flat-top area and equals to the pulse voltage. Then, we have
\begin{equation}\label{eq:VSTL1}
    V_0 = \Delta B \dfrac{\delta S}{\delta t}
        = \Delta B \dfrac{c(r_o-r_i)}{\sqrt{\varepsilon_f\mu_f}}.
\end{equation}
For given core material, $\sqrt{\varepsilon_f\mu_f}$ and the maximum value of $\Delta B$ are determined. Hence, the outer and inner radius difference $(r_o-r_i)$ must be larger enough to support needed $V_0$.
Substitute Eq.(\ref{eq:VSlength2}) into Eq.(\ref{eq:VSTL1}), we get
\begin{equation}\label{eq:VSTL2}
    V_0\tau \leqslant 2\Delta B(r_o-r_i)L.
\end{equation}
or:
\begin{equation}\label{eq:VSTL3}
    V_0\tau \leqslant 2\Delta BS.
\end{equation}
In the derivation Eq.(\ref{eq:VSlength1}), we included the reflection process of the high voltage pulse from the shorted end. The reflected pulse has the same current as the incident pulse. This means that the maximum $\Delta B$ we can utilize in Eq.(\ref{eq:VSTL3}) is only half of $\Delta B_m$ in Eq.(\ref{eq:VS4}) since we have to reserve the other half for the reflected pulse, otherwise, the cores will become saturated when reflected pulse comes back. After substitute $\Delta B = \Delta B_m/2$ into Eq.(\ref{eq:VSTL3}), we get Eq.(\ref{eq:VS4}) again. And we know from above derivation that $L$ and $(r_o-r_i)$ are determined by pulse length and voltage amplitude, respectively. The combination of these two conditions give the volt-second criteria.

\subsection{Wave's Reflection inside the Cell vs Droop}
When the electrical length of the core cell is much shorter than that of the voltage pulse, the magnetic core can be considered as a large inductance. Even with a perfectly flat top input voltage pulse, the voltage shape at the gap exhibit a slow dropping characteristic, the droop.
\begin{equation}\label{eq:Droop}
    V(t) = 2 V_0 \dfrac{R_b\parallel R_c}{R_b\parallel R_c+Z_0}e^{-t/\tau}.
\end{equation}
Here we assume the Blumlein line has a characteristic impedance $Z_0$ and is charged to a voltage $2V_0$, $R_b$ represents the beam resistance in parallel with any compensating resistor $R_c$, $\tau=L/Z_{eff}$ is the decay time constant of the circuit. $L$ is the inductance, $Z_{eff}$ is the effective impedance of the line in parallel with $R_b$. The voltage droop can be kept small by keeping $L$ large or by keeping $Z_{eff}$ low enough.

In transmission line type modules, the compensating resistor $R_c$ is arranged in such a way that $R_b\parallel R_c\parallel R_m$ is matched with Blumlein line impedance $Z_0$, where $R_m$ is the characteristic impedance of the core cell. Under such a condition, there will be no reflection when high voltage pulse travels into the core cell. In the core cell, when the pulse is reflected back from the shorted end and arrive at the gap region, the impedance is not matched anymore looking from the core cell into the power supply cable and $R_b\parallel R_c$. Hence, most of the pulse will be reflected again at the gap interface, and the pulse will travel back and forth inside the core cell. Each time it arrives at the gap interface, a small part transmits into the gap and the power supply cable and cause the gap voltage drops a little. When the electrical length of the core cell gets shorter and shorter, this stepwise style gap voltage dropping degenerates into the droop described in  inductance type modules.

\subsection{Wave Velocity in The Core Cell}
In the transmission line type module, the long enough process of pulse transportation inside the core cell is the key point to avoid the gap voltage droop. Hence, the accurate wave velocity determination in module design and experiment is important. In the core cell, the gaps between the ferrite and conductors change the wave velocity, as illustrated in Fig.~\ref{Ferrite_Transmission_Line}.
\begin{center}
\includegraphics[width=8cm, angle=0]{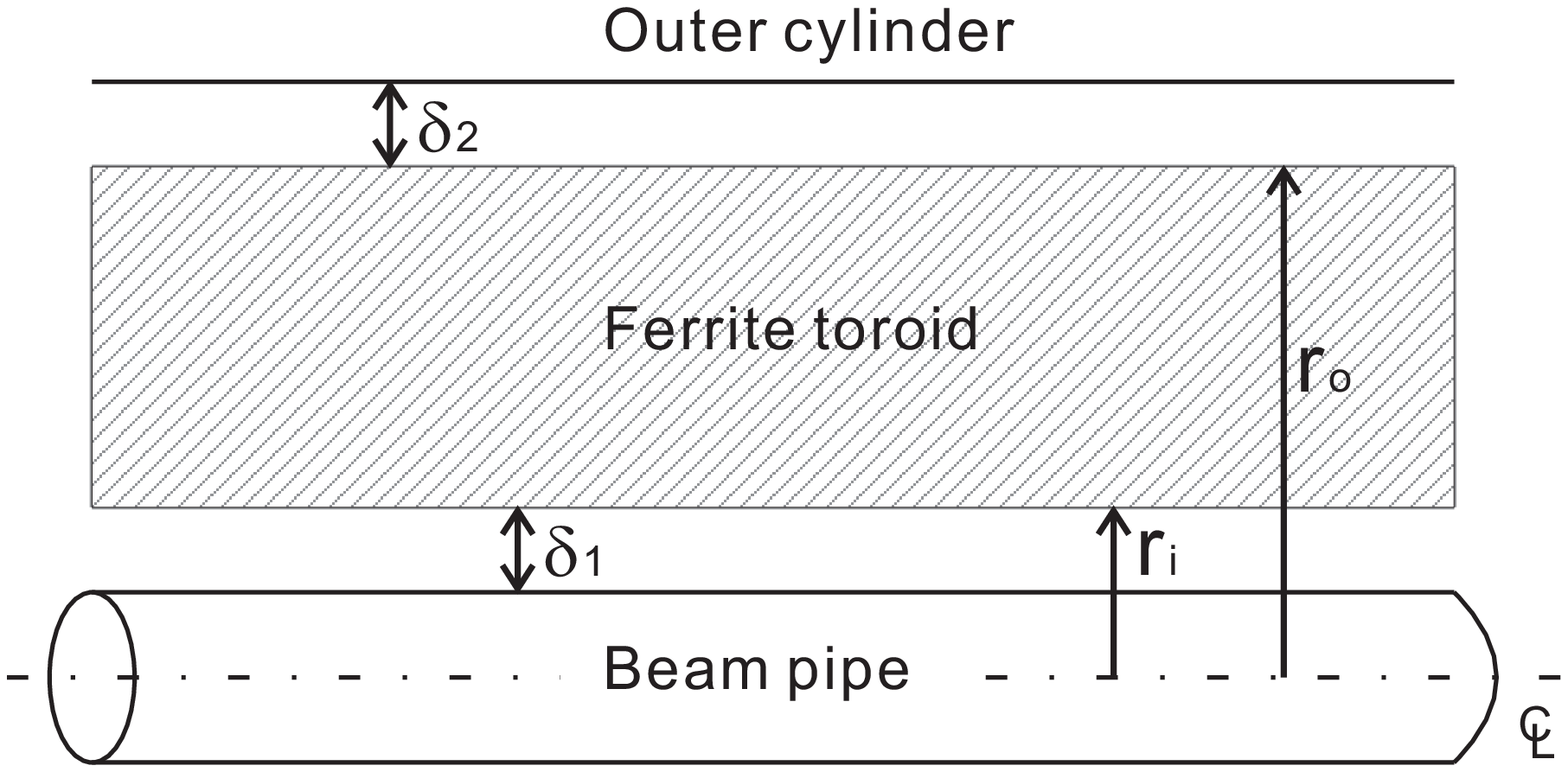}
\figcaption{\label{Ferrite_Transmission_Line}Ferrite transmission line.}
\end{center}

From the electric field distribution inside the core cell,
\begin{eqnarray}\nonumber
    V&=&\int_{r_i-\delta_1}^{r_i}Edr + \int_{r_i}^{r_o}Edr +
        \int_{r_o}^{r_o+\delta_2}Edr \\ \nonumber
     &=&\dfrac{q}{2\pi\varepsilon_0}
       (\dfrac{1}{\varepsilon_g}\ln\dfrac{r_i}{r_i-\delta_1}+
        \dfrac{1}{\varepsilon_f}\ln\dfrac{r_o}{r_i}+ \\
     & &\dfrac{1}{\varepsilon_g}\ln\dfrac{r_o+\delta_2}{r_o} ).
\end{eqnarray}
The capacitance of unit length of the transmission line can be calculated:
\begin{eqnarray}\nonumber
    C_f&=&\dfrac{2\pi\varepsilon_0}
         {\dfrac{1}{\varepsilon_g}\ln\dfrac{r_i(r_o+\delta_2)}{r_o(r_i-\delta_1)}
         +\dfrac{1}{\varepsilon_f}\ln\dfrac{r_o}{r_i} } \\
       &=&\dfrac{2\pi\varepsilon_0\varepsilon_e}{\ln\dfrac{r_o+\delta_2}{r_i-\delta_1}}.
\end{eqnarray}
 where,
\begin{equation}
    \varepsilon_e = \dfrac{\varepsilon_g\varepsilon_f\ln\dfrac{r_o+\delta_2}{r_i-\delta_1}}
         {{\varepsilon_f}\ln\dfrac{r_i(r_o+\delta_2)}{r_o(r_i-\delta_1)}+
          \varepsilon_g\ln\dfrac{r_o}{r_i}}.
\end{equation}
And, in a similar way,
\begin{eqnarray}\nonumber
    L_f i &=&\int_{r_i-\delta_1}^{r_o+\delta_2}B dr \\ \nonumber
          &=&\int_{r_i-\delta_1}^{r_i}\mu_0\dfrac{i}{2\pi r} dr +
            \int_{r_i}^{r_o}\mu_0\mu_f\dfrac{i}{2\pi r} dr + \\
          & &\int_{r_o}^{r_o+\delta_2}\mu_0\dfrac{i}{2\pi r} dr.
\end{eqnarray}

The inductance of unit length of the transmission line is:
\begin{eqnarray} \nonumber
    L_f &=&\dfrac{\mu_0}{2\pi}(\ln\dfrac{r_i (r_o+\delta_2 )}{r_o(r_i-\delta_1 )}
          +\mu_f\ln\dfrac{r_o}{r_i}) \\
        &=&\dfrac{\mu_0\mu_e}{2\pi}\ln\dfrac{r_o+\delta_2}{r_i-\delta_1}.
\end{eqnarray}
Where
\begin{equation}
    \mu_e = \dfrac{\ln\dfrac{r_i (r_o+\delta_2 )}{r_o(r_i-\delta_1 )}
           +\mu_f\ln\dfrac{r_o}{r_i}}{\ln\dfrac{r_o+\delta_2}{r_i-\delta_1}}.
\end{equation}
The phase velocity in the core cell transmission line is
\begin{eqnarray}\nonumber
    v_p &=&\dfrac{1}{\sqrt{L_f C_f}} \\ 
          &=&c\sqrt{\dfrac{
          {\dfrac{1}{\varepsilon_g}}\ln\dfrac{1+\delta_2/r_o }{1-\delta_1/r_i}+
           \dfrac{1}{\varepsilon_f}\ln\dfrac{r_o}{r_i}}
          {\ln\dfrac{1+\delta_2/r_o } {1-\delta_1/r_i}+\mu_f\ln\dfrac{r_o}{r_i}}}.
\end{eqnarray}

For a test module, the inner and outer radius of the ferrite toroid are 136 mm and 446.5 mm, respectively. $\delta_1$ and $\delta_2$ both are 1.5 mm. The pulse speed is 8 percent faster than zero $\delta_1$ and $\delta_2$ situation if we choose $\varepsilon_f = 13$ and $\mu_f = 400$.

\section{Summary}

Both transformer model and transmission line model are good in explaining there corresponding LIA acceleration modules. The differences between these two types of LIA modules are distinguished and clarified by comparing the structure and acceleration mechanisms. Some important concepts are re-examined based on the transmission line model.

\acknowledgments{The authors thank members of the accelerator group of CAEP/IFP for fruitful comments and discussions.}

\vspace{3mm}

\bibliographystyle{apsrev}
\bibliography{IC_V9}

\end{multicols}

\clearpage

\end{CJK*}
\end{document}